# Pump-Enhanced Continuous-Wave Magnetometry Using Nitrogen-Vacancy Ensembles

Sepehr Ahmadi, Haitham A. R. El-Ella,[*] Jørn O. B. Hansen, Alexander Huck, and Ulrik L. Andersen

*Department of Physics, Technical University of Denmark, 2800 Kongens Lyngby, Denmark*

Ensembles of nitrogen-vacancy centers in diamond are a highly promising platform for high-sensitivity magnetometry, whose efficacy is often based on efficiently generating and monitoring magnetic-field-dependent infrared fluorescence. Here, we report on an increased sensing efficiency with the use of a 532-nm resonant confocal cavity and a microwave resonator antenna for measuring the local magnetic noise density using the intrinsic nitrogen-vacancy concentration of a chemical-vapor deposited single-crystal diamond. We measure a near-shot-noise-limited magnetic noise floor of 200 pT/$\sqrt{\text{Hz}}$ spanning a bandwidth up to 159 Hz, and an extracted sensitivity of approximately 3 nT/$\sqrt{\text{Hz}}$, with further enhancement limited by the noise floor of the lock-in amplifier and the laser damage threshold of the optical components. Exploration of the microwave and optical pump-rate parameter space demonstrates a linewidth-narrowing regime reached by virtue of using the optical cavity, allowing an enhanced sensitivity to be achieved, despite an unoptimized collection efficiency of < 2%, and a low nitrogen-vacancy concentration of about 0.2 ppb.

## I. INTRODUCTION

The nitrogen-vacancy (N-$V$) center, an atomlike defect within a diamond crystal lattice, is an auspicious quantum sensor because of its readily polarized and detected spin state [1]. The characteristics of the N-$V$ center combine high sensitivity and spatial resolution allowing for the detection of spatial temperature gradients [2,3], electric fields [4], and magnetic fields [5–7]—all at room temperature—by measuring the fluorescence contrast via optically detected magnetic resonance (ODMR). By using an $N$-size ensemble of N-$V$ centers, the collective sensitivity is improved by a factor of $N^{-1/2}$ [8]. Developing diamond ensemble-based magnetic sensors presents advantages over their atomic-vapor and superconducting equivalents, particularly in terms of their relative simplicity, integrability in a variety of devices, and biological compatibility [9]. The methods developed for the collective control and readout of a N-$V$ ensemble for sensing applications should also be useful for scalable quantum-information schemes using ordered ensembles [10], or for potentially coupling N-$V$ ensembles to atomic vapors to create hybrid quantum systems for fundamental and applied experiments [11].

In this article, we present a proof-of-principle system for continuous-wave N-$V$-ensemble-based magnetometry which combines an optical and a microwave resonator. In combination with lock-in amplification and the simultaneous excitation of all three hyperfine lines, a near-shot-noise-limited noise floor of approximately 200 pT/$\sqrt{\text{Hz}}$ is achieved from the inherent $^{14}$N-$V^{-}$ concentration of an off-the-shelf single-crystal diamond (approximately 0.2 ppb) while collecting an estimated < 2% of the generated fluorescence. To put these parameters in context, assuming that sensitivity unquestionably scales with $N^{-1/2}$ and the detection efficiency, the work presented here should be compared to, for example, Refs. [12,13], which report measured sensitivities in the roughly 15- and 290-pT/$\sqrt{\text{Hz}}$ ranges, respectively, using highly engineered diamonds with enhanced N-$V$ concentrations > 100 ppb and a detection efficiency that is > 15%. The system presented in this article shows comparable sensitivity despite the lower concentration and collection efficiency, and it is anticipated to provide a solid foundation for developing further optimized and miniaturized cavity-based devices. These devices are to be implemented and combined in a variety of sensing applications, most notably in scenarios involving live biological tissue. Other works which report pT-fT sensitivities using N-$V$ ensembles include that of Refs. [14,15]; however, these reported sensitivities are limited to ac fields, as they are achieved using spin-echo pulse sequences—unlike the continuous-wave approach presented in this work, which is applicable to both dc and ac fields.

The first section of this article briefly overviews the motivation of this work, followed by an overview of the N-$V$-center electronic structure, along with the experimental methods used. This overview is followed by a discussion of the optical-cavity parameters and a brief assessment of the system's total optical loss, leading to an estimate of the number of N-$V$ centers collectively excited and the proportionality between the intracavity power and the ensemble excitation rate. Next, the ODMR spectrum and its equivalent lock-in spectrum are presented, and the characteristics of the optimal parameters reachable with our current configuration are discussed. The measured spectra are compared with a theoretical model which highlights the occurrence of linewidth narrowing and allows for the derivation of an optimum shot-noise-limited sensitivity. Finally, measurements of the magnetic noise spectral density and an external magnetic field using the optimized system are presented.

[*]haitham.el@fysik.dtu.dk

## II. EXPERIMENTAL METHOD AND SETUP

Obtaining high densities and efficiently employing them in a sensing scheme poses complications in terms of collective control and signal readout. Specifically, the inherent difficulties with increased densities spanning larger volumes involve insufficient collective polarization, inhomogeneous broadening, and nonoptimal signal collection.

Large densities are usually obtained through high-temperature high-pressure growth or extended nitrogen ion implantation and annealing, both of which suffer from poor N-to-N-$V$ conversion efficiencies (< 20% [16]) and lead to a large concentration of N nuclear spins (in the form of single and aggregate substitutional defects) alongside $^{13}$C nuclear spins. These nuclear spins inhomogeneously broaden the collective N-$V$ spin resonance, and while such broadening can be mitigated by using $^{12}$C purified diamond [17], the problem of low N-$V$ creation efficiencies accompanying degradation of the collective coherence and crystal integrity still remains. At the same time, efficient collection of the generated fluorescence involves circumventing the limitations of the high refractive index of diamond, which results in a small critical angle of escape and significant light loss through total internal reflection. Different approaches have been studied to overcome this problem, involving, e.g., applying a solid immersion lens [18,19], coupling N-$V$ centers to photonic crystals [20], using a dielectric antenna [21] or side-collection detection [14]. Alternative strategies intended to circumvent these issues have also been explored through sensing based on cavity-assisted absorption of the shelving state [22,23], and amplifying the ODMR fluorescence by trapping pump light within the diamond through total internal reflection [13].

In general, single and ensemble N-$V$-based continuous-wave sensing is limited by the ratio of the spectral ODMR linewidth to the fluorescence contrast. For a given ensemble, maximizing this ratio requires spatially uniform spin polarization and readout pump rates across the ensemble volume, which allows a saturated fluorescence regime to be reached. Reaching a saturated regime becomes more difficult with increasing ensemble volume and density, given (a) the negligible absorption cross section of single N-$V$ centers, (b) the spatial inhomogeneity of the nuclear spin bath, (c) the generated spin noise resulting from spatially inhomogeneous fields which can only identically polarize a fraction of the ensemble, and (d) the inherent power broadening that accompanies increased polarization rates to maximize ODMR fluorescence contrast. One method of collectively tackling these issues is through addressing the spatial uniformity and efficiency of the spin polarization and readout pump fields with respect to a given ensemble volume, while ensuring that the readout pump rate is large enough to counteract power broadening [24]. In this work, these issues are addressed by using an optical cavity resonant with the readout pump light, in conjunction with a microwave split-ring resonator antenna. The antenna allows for a spin-polarizing field to be applied uniformly across the ensemble volume that is excited by the readout pump, while the cavity amplifies the power of the input beam to collectively excite the ensemble at a fast-enough rate to reach a linewidth-narrowing regime.

The electronic configuration of the N-$V$ center [a simplified schematic is shown in Fig. 1(a)] exhibits spin-dependent fluorescence by virtue of the difference between the $|m_s = \pm 1\rangle$ and the $|m_s = 0\rangle$ spin-level intersystem crossing rates ($k_I$) from the $^3E$ excited triplet state to the shelving singlet states ($^1A_1 \leftrightarrow {}^1E$). In a simplified continuous-wave (CW) picture for an ensemble of identical systems, the radiative relaxation rate ($k_r$) of all of the excited-state levels competes with $k_I$ to bring about a fluorescence contrast ($\mathcal{C}$) based on the initial distribution of the spin-state population in the triplet ground states ($^3A_2$). This ground-state spin population is set by the ratio between the optical readout pump rate ($\Gamma_p$) from above-band excitation (in this case, a 532-nm laser), and the resonant microwave (MW) drive rate of the spin levels (Rabi frequency, $\Omega$), of which the optical above-band excitation mechanism is near-perfectly spin conserving. The presence of a local magnetic field breaks the degeneracy of the $|m_s = \pm 1\rangle$ spin levels in the ground and excited states, splitting them with a scaling proportional to the product of the gyromagnetic ratio of $\gamma_e \sim$ 28 MHz/mT and the projection angle of the field along the N-$V$ symmetry axis. The four possible crystallographic orientations of a single N-$V$ within the diamond unit cell (along all four $\langle 111 \rangle$ directions) result in four distinct N-$V$ group alignments for a given magnetic field. In the experimental setup, the splitting of these groups is tuned via the placement of a permanent rare-earth magnet on a three-axis translation stage in the vicinity. A coil magnet is also placed next to the diamond (offset from the optical-cavity path) to generate weaker fields for assessing the system's response to low-frequency ac magnetic fields. The presence and magnitude of a local magnetic field is directly detected by monitoring the change in fluorescence of one of these aligned groups as a function of the applied MW frequency $\omega_c/2\pi$. As a result, the detection sensitivity using the ODMR spectrum $\mathcal{S}_{\mathrm{CW}}$ is limited by the inverse of the maximum MW-dependent rate of fluorescence change, $[\max\{d_\omega \mathcal{S}_{\mathrm{CW}}\}]^{-1}$, where $d_\omega$ denotes the derivative with respect to the MW frequency.

Confocal cavity-assisted ODMR is carried out using the native $^{14}$N-$V^-$ concentration of a polished and untreated single-crystal diamond (6 × 6 × 1.2 mm, Element Six) grown using chemical vapor deposition (CVD), with a quoted substitutional nitrogen concentration [$N_s$] of < 1 ppm. The diamond is mounted on a homebuilt apertured circuit board, patterned with a MW split-ring resonator and, as shown in the schematic in Fig. 1(b), placed vertically between two confocal-cavity mirrors at a Brewster angle of $\theta_B = 67° \pm 0.4°$ relative to the cavity's longitudinal axis. Both cavity mirrors, $R_1$ and $R_2$, have a curvature radius of 10 cm and are antireflection coated for 532 nm on the flat ends. The measured reflectivities are $R_1 = 94.8\% \pm 0.1\%$ and $R_2 = 99.8\% \pm 0.1\%$, giving a projected finesse of $113 \pm 4$. The measured finesse of the empty cavity is $\mathcal{F} = 114$, shown in Fig. 2(a), agrees well within the error of the projected value. An unequal mirror reflectivity is chosen in order to approach impedance matching conditions when incorporating the diamond. The cavity is pumped with a phase-modulated and $p$-polarized 532-nm laser possessing a single longitudinal mode (Verdi SLM Coherent, rms < 0.03% from 10 Hz–100 MHz), and Pound-Drever-Hall (PDH) locked using the transmitted light through $R_2$, and a piezoelectric actuator attached to the $R_1$ mirror mount. The confocal configuration of these mirrors results in a Laguerre-Gaussian LG$_{00}$ mode with a $1/e$ beam waist of 92 $\mu$m and a Rayleigh length of approximately 50 mm. With the diamond incorporated at its $\theta_B$ angle, and accounting for the standing-wave spatial intensity profile, an effective $1/e$ excitation volume of about $3.5 \times 10^{-2}$ mm$^3$ is obtained, considering the LG$_{00}$ transverse beam profile. The overall influence of the spatial standing-wave intensity variations are observed to be negligible when performing comparative measurements with and without the use of the cavity mirrors for similar optical excitation powers. Owing to the expected uniform spatial distribution of the N-V centers, the same total fluorescence rate is expected when comparing a flat and sin$^2$ spatial profile with the same average intensity.

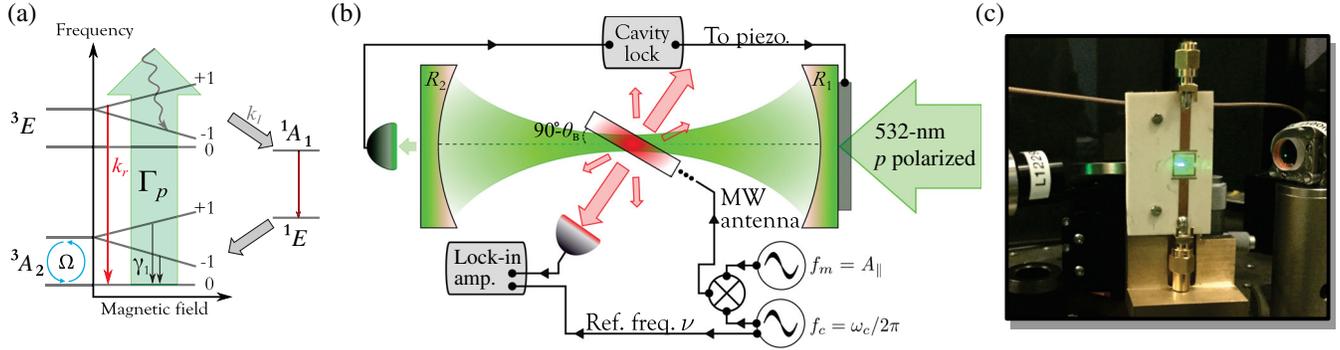

FIG. 1. (a) A simplified diagram of the N-V energy levels, highlighting the most significant decay and excitation pathways described in the text, as well as the splitting of the ground- and excited-state spin levels as a function of external magnetic field strength. (b) Simplified schematic of the experimental setup highlighting the most essential components used to measure ODMR from the diamond. The diamond is placed at a Brewster angle $\theta_B$ relative to the cavity axis, which is defined by two mirrors with reflectivities $R_1$ and $R_2$, where the input mirror reflectivity $R_1 < R_2$. The cavity is pumped through the input mirror with phase-modulated $p$-polarized light and is locked using the PDH technique with a servomechanism that monitors the transmitted light through $R_2$ and actuates a piezoelectric stack attached to $R_1$. The N-V resonance is driven by a modulated frequency $f_c = \omega_c/2\pi$ that is mixed with $f_m = A_\parallel = 2.16$ MHz generated from a second signal generator. An antenna attached to the apertured diamond holder is used to deliver the mixed drive frequency. A modulation reference signal at frequency $\nu$ is sent to a lock-in amplifier which is connected to the detector measuring the ODMR fluorescence generated from the diamond. (c) A picture of the diamond mounted on the apertured MW antenna, placed between the cavity mirrors.

For sensing-based measurements, fluorescence is collected directly from the large face of the diamond using either a NA = 0.7 objective (Mitutoyo, with an approximate transmission efficiency of 77% at 700 nm), or a NA = 0.79 condenser lens (Thorlabs ACL25416U-B, with an approximate transmission efficiency of 99.7% at 700 nm) and filtered using a long-pass filter with a 600-nm cut-on wavelength. Considering the refractive index of diamond, the numerical aperture of the objective, and the loss from the remaining optical components, the total collection efficiency is estimated to be < 2%. The collected fluorescence is focused onto a Si-biased detector (Thorlabs DET36A with an approximate quantum efficiency of 70% at 700 nm) attached with a 10 k$\Omega$ load resulting in a 400-kHz detection bandwidth. The detected signal is noise filtered and amplified using a lock-in amplifier (Stanford Research Systems SR510) through either amplitude or sine-wave frequency modulation of the MW drive (30-kHz modulation rate and 0.5-MHz modulation depth), for which the in-phase quadrature is output. The lock-in amplifier input is set with a bandpass filter centered at the modulation frequency (6-kHz bandwidth), and the output is set with a 1-ms time constant, thereby imposing a first-order low-pass filter with a corner frequency at 159 Hz. For simultaneously driving all three $^{14}$N hyperfine lines, the modulated MW is mixed with a 2.16-MHz drive (corresponding to the axial hyperfine splitting frequency $A_\parallel$). The MW drive frequency $\omega_c/2\pi$ is delivered using a split-ring resonator which has a measured resonance at 2.884 GHz and a bandwidth of about 91 MHz. The spatial uniformity of the delivered MW power is deduced from spatially dependent measurements of the Rabi frequency $\Omega$ using confocal microscopy, which is observed to vary by < 2.5% over a region of about 1 mm$^2$, across the 1.2-mm thickness of the diamond. The equivalent $\Omega$ of the applied MW power is determined using pulsed excitation in the cavity configuration.

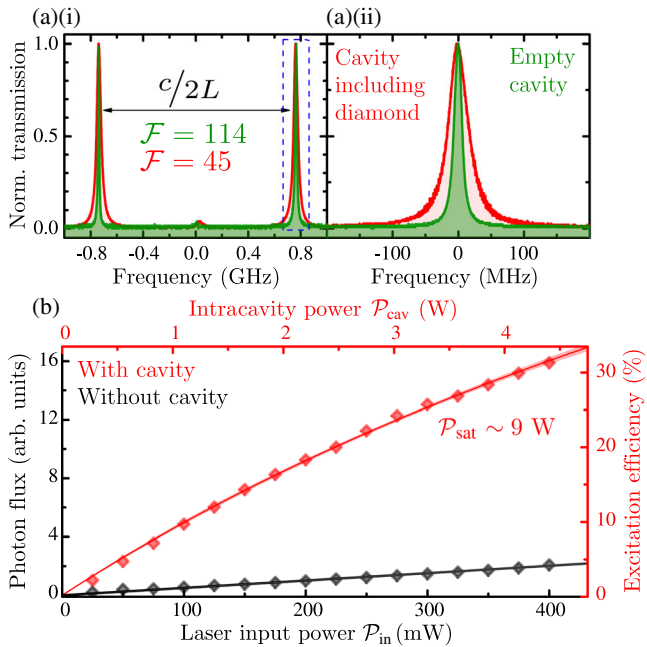

FIG. 2. (a)(i) Transmission spectrum of the cavity as a function of the laser frequency detuning for the cavity without (green) and with (red) the diamond, showing the dominance of the $LG_{00}$ mode and a small negligible peak of the $LG_{01}$ mode (evident from the displacement by half of the free spectral range) for both instances. (a)(ii) Enlargement of one of the transmission peaks highlighting the linewidth increase from approximately 6.5 to 17 MHz. (b) Relative power-dependent photon flux in arbitrary units for the collected light from the diamond without (black points) and with (red points) the cavity, plotted as a function of the cavity input $\mathcal{P}_{in}$ and intracavity powers $\mathcal{P}_{cav}$. The excitation efficiency $\mathcal{R}/\mathcal{R}_{sat}$ is also shown on the right scale for the cavity-enhanced photon flux. The black trace is a linear fit, while the red trace is a power-law fit, as discussed in the text.

### III. CAVITY-ENHANCED ODMR READOUT

#### A. Characteristics of the confocal cavity

The enhancement of the input laser power by the confocal cavity is estimated directly from the measured finesse, $\mathcal{F}$. With the inclusion of the diamond and its associated absorption losses, a 60.5% reduction of the finesse is observed down to $\mathcal{F} = 45$, as shown in Fig. 2(a), along with a full free spectral range scan highlighting the dominance of the $LG_{00}$ mode. The on-resonance relationship between the intracavity power $\mathcal{P}_{cav}$ and the input power $\mathcal{P}_{in}$ is $\mathcal{P}_{cav} \simeq \mathcal{P}_{in}T_1/(1-\varrho)^2$, where $T_1$ is the transmission of the input mirror and $\varrho = \sqrt{R_1R_2e^{-\alpha}}$ is the cumulative round-trip loss product, with $\alpha$ being the propagation-loss coefficient. $\varrho$ is calculated from the polynomial root of its relation with $\mathcal{F}$ through $\mathcal{F}\varrho + \pi\sqrt{\varrho} - \mathcal{F} = 0$ and, in the absence of the diamond, is assumed to be a function solely dependent on the product of $R_1R_2$. Using this relationship, a measured power-dependent flux rate is plotted in Fig. 2(b) as a function of $\mathcal{P}_{cav}$ and $\mathcal{P}_{in}$. A clear transition from a linear to a nonlinear dependence is observed which is fitted with $\mathcal{R} = \mathcal{R}_{sat}\mathcal{P}/(\mathcal{P} + \mathcal{P}_{sat})$ to estimate the excitation efficiency $\mathcal{R}/\mathcal{R}_{sat}$ and the projected saturation power $\mathcal{P}_{sat}$. The excitation efficiency is estimated to reach about 31% for $\mathcal{P}_{in} \sim 0.4$ W and 50% for $\mathcal{P}_{in} \sim 0.87$ W, for which the latter corresponds to an intracavity power of $\mathcal{P}_{sat} \sim 9$ W.

By measuring the finesse of the empty cavity, the product of the two mirror reflectivities, $R_1R_2$, is determined with adequate precision. This determination of $R_1R_2$ from an empty cavity allows for the reasonable assumption that the product $\alpha$ from the measured finesse of the loaded cavity is attributable solely to the diamond. This loss is composed of diamond bulk- and surface-based absorption, $\alpha_{ab}\ell$ and $\alpha_{surf}$, birefringence-related loss $\alpha_{br}$, and scattering losses $\alpha_{sc}$ due to surface-roughness-based scattering for every round-trip: $\alpha = \alpha_{ab}\ell + 4\alpha_{surf} + 4\alpha_{sc} + \alpha_{br}\ell$. Given the magnitudes of power, the laser wavelength, and the 600-nm cut-on wavelength of the bandpass filter, loss associated with the discharging of $^{14}$N-$V^- \rightarrow ^{14}$N-$V^0$ is considered to be negligible [25]. To decompose $\alpha$, the amount of green light reflected from the surface of the diamond while the cavity is locked is measured. The total fraction of this reflected power to intracavity power corresponds to about 0.006, of which approximately 80% is $s$-polarized light. Using this measured reflection-based loss, and assuming that surface absorption is negligible (as the surfaces are cleaned with acid and constitute a negligible fraction of the beam path), an absorption-loss coefficient of $\alpha_{ab} \sim 0.0301$ mm$^{-1}$ is estimated for $\ell = 2 \times 1.3$ mm.

Apart from the absorption of N-$V^-$, absorption around 532 nm can also be attributed principally to the nitrogen-vacancy-hydrogen (N-$V$-H) defect inherent in CVD-grown diamond [26]. Using the reported absorption coefficients in Ref. [26] for similar CVD diamonds ($\alpha_{N-V^0-H} \sim 0.03$ mm$^{-1}$), along with the absorption cross section of N-$V$ reported in Ref. [27] ($\sigma_{N-V} = 3.1 \times 10^{-15}$ mm$^2$), the [N-$V^-$] concentration can be approximated through $\alpha_{ab} \approx \sigma_{N-V}[N-V^-] + \alpha_{N-V^0-H}$. An estimated concentration of about $6 \times 10^{10}$ mm$^{-3}$ is achieved, which translates to approximately 0.2 ppb $^{14}$N-$V^-$ (about 0.02% of the quoted [$N_s$] upper-bound concentration). These estimated concentrations are corroborated with estimates obtained through fluorescence count-rate measurements which indicate a concentration of about 0.16 ppb. Altogether, an order of magnitude estimate of $10^9$ N-$V^-$ within the excitation volume is determined. Although the inability to exactly discern competing absorption species and their mechanisms results in a large uncertainty in using these assumptions, the estimate represents a realistic order of magnitude which reflects values reported in the literature for similar CVD diamonds, and from which the proportionality between the intracavity pump power and the excitation rate can be estimated to be on the order of $\varepsilon \sim 0.3$ kHz/mW.

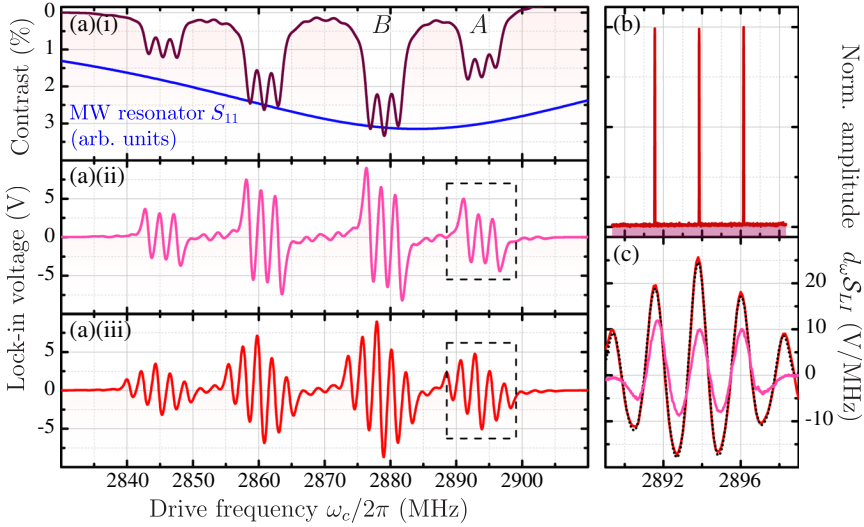

FIG. 3. (a)(i) Amplitude-modulated lock-in spectrum (purple trace) plotted in units of measured signal contrast as a function of MW drive frequency, with the frequency-dependent reflection parameter ($S_{11}$) of the split-ring resonator antenna (blue trace) incorporated for comparison (not sharing the y axis with the red trace). The frequency-modulated spectrum using single-frequency excitation (a)(ii) and three-frequency excitation (a)(iii) are shown for comparison. (b) Measured three-frequency excitation spectrum (a wide-span measurement is shown in the Supplemental Material [28]). (c) Plot of the derivative $d_\omega \mathcal{S}_{LI}$ of peak $A$ outlined in (a)(ii) and (a)(iii) using the same color. A simulation of the signal using Eq. (4) is overlaid with the measured trace in (a)(iii) (the black dotted line).

## B. Lock-in amplified magnetic resonance detection

The full ODMR spectrum of the $^3A_2$ spin states and the corresponding lock-in detected signals are shown in Fig. 3. The spectrum in Fig. 3(a)(i) is obtained through lock-in amplification using a pulsed MW signal, and highlights the deliberate alignment of the external magnetic field along one of the $\langle 111 \rangle$ crystallographic directions, as well as the influence of the split-ring resonator bandwidth. Given the four possible crystallographic orientations of the N-V defect, perfect alignment with only one axis orientation at a time is possible for a simple magnetic field, which brings about the spin-level degeneracy of the three other orientations. For an ensemble, splitting of the lines is generally desired in order to circumvent the degraded resonance line shape resulting from residual stray fields that break the degeneracy of the four subgroups. It is also necessary in the case of pulsed schemes where optimal coherence times are required [29], but these advantages are brought about at the expense of contrast, which is clearly visible in Fig. 3(a)(i) when comparing the leftmost peak to the two inner peaks. In the case of CW measurements and sensing schemes, such splitting is not strictly required and the advantage in detection of the enhanced contrast of three degenerate lines outweighs the reduced collective coherence. However, the three subgroups' sensitivity is degraded by the nonparallel projection angle of an external magnetic field. Irrespective of how the sensed field is aligned with respect to the three subgroups, the gyromagnetic ratio is—at best—reduced by a factor of $\cos(54.7°)$.

The amplitude ratios of the outer and inner peaks, designated here as group $A$ and group $B$, respectively [labeled in Fig. 3(a)], do not possess a 1:3 ratio in this case due to the use of the split-ring resonator, which ensures more spatially uniform power delivery at the expense of power uniformity over the desired ODMR frequency window. The $S_{11}$ trace of the loaded MW resonator is overlaid in Fig. 3(a)(i) (in normalized arbitrary-loss units) and shows a peak at around 2.884 GHz, which is centered between the $|m_s = +1\rangle$ spin levels of groups $A$ and $B$.

As introduced previously, the sensitivity of the resonances to the presence of fluctuating fields is directly proportional to $[\max\{d_\omega \mathcal{S}_{CW}\}]^{-1}$. This product can be enhanced by modulating the MW field and amplifying the resulting ac signal using a lock-in detector, resulting in a frequency-modulated ODMR signal $\mathcal{S}_{LI}$, which is proportional to the derivative of the unmodulated signal and is shown in Fig. 3(a)(ii). The resulting spectra displays features with slopes that are approximately $2\times$ steeper, and presents a higher field sensitivity at the expense of the lock-in-imposed bandwidth. However, due to the presence of three hyperfine levels from the $^{14}$N nuclear spin and their considerable spectral overlap, single-frequency excitation results in an unoptimized contrast—and therefore a reduced $\max\{d_\omega \mathcal{S}_{LI}\}$—relative to the optimal contrast expected from a single ideal spin resonance. The reduced contrast may be mitigated by either working at the excited-state level anti-crossing regime where all N nuclei are polarized upon green-light excitation [30], or by simultaneously exciting all three frequencies, which should result in the same enhanced contrast, using the measured MW excitation frequency spectrum shown in Fig. 3(b). The latter approach is more practical, as it avoids the restriction of working at a set magnetic field strength and alignment angle [30,31]. Simultaneous excitation of all three hyperfine transitions is carried out by mixing the modulated driving frequency with a sine wave oscillating at a frequency equal to the axial hyperfine constant $A_\parallel = 2.16$ MHz (the splitting of the three hyperfine lines), which results in the spectrum shown in Fig. 3(a)(iii). The resulting slopes in Fig. 3(a)(iii) are approximately 2–2.5 times the slopes generated from the single-frequency modulated signal, as can be seen in Fig. 3(c), with further enhancement dependent on the optimization of the modulation depth, and the $\Gamma_p/\Omega$ ratio which judges the degree of linewidth narrowing brought about.

Both CW and lock-in spectra can be simulated using the steady-state solution of a five-level Bloch equation [32], set up with the rates reported in Ref. [33]. An analytical expression is defined in terms of the steady-state solutions of the excited-state populations $\mathcal{I}_{\mathrm{CW}}$ as a function of $\Gamma_p$, $\Omega$, and the detuning of the drive frequency $\omega_c/2\pi$ relative to the peak resonance frequency $\omega_0/2\pi$ [28]. For the three hyperfine lines, the expression can be defined as a sum of three Lorentzian distributions:

$$\mathcal{S}_{\mathrm{CW}}(\omega_c) = \sum_{m_I=\{-1,0,1\}} V_0 \left( \frac{(\mathcal{I}_{\mathrm{CW}}^{\Omega=0} - \mathcal{I}_{\mathrm{CW}}^{\Delta=0})\gamma^2}{\Delta^2 + \gamma^2} \right), \quad (1)$$

where $V_0$ is the off-resonance detected voltage, $m_I$ is the nuclear spin quantum number, $\gamma$ is the half width at half maximum, and the detuning is expressed through $\Delta = (\omega_c - \omega_0) + 2\pi m_I A_\parallel$. This expression includes the influence of both power broadening and linewidth narrowing through $\mathcal{I}_{\mathrm{CW}}$, providing a basis to define the lock-in equivalent signal, which is a product of a cosine reference signal $\mathcal{S}_{\mathrm{ref}} = \cos(2\pi \nu t)$ and a detected signal, both modulated at a reference frequency $\nu$. The modulation of $\omega_c/2\pi$ is carried out at the same frequency $\nu$ with a modulation depth $m$ using a sine-wave function, resulting in a modulation index $\beta = m/\nu$. The ODMR spectrum $\mathcal{S}_{\mathrm{CW}}$ becomes a function of

$$\omega(t) = \sum_{n=-\infty}^{\infty} J_n(\beta) \sin(\omega_c t + n\nu 2\pi t), \quad (2)$$

where $J_n$ is a Bessel function of the first kind of order $n$ [28]. Assuming that the reference and modulated signals are perfectly in phase, the expression for the lock-in signal takes the form of

$$\mathcal{S}_{\mathrm{LI}}(\omega_c) = \mathcal{A} V_0 \mathcal{S}_{\mathrm{CW}}(\omega(t)) \mathcal{S}_{\mathrm{ref}}, \quad (3)$$

where a lock-in voltage prefactor $\mathcal{A} \propto 10g$ is proportional to the lock-in amplifier gain factor $g$. For $\beta > 1$, the power spectrum of Eq. (2) shows approximately $n \simeq \lceil \beta \rceil$ significant peaks which possess an amplitude that is at least 1% of the central unmodulated carrier frequency [34]. Using this picture, expression (3) becomes equivalent to the sum of two sets of $n/2$ out-of-phase spectra, each driven by an $\Omega$ scaled by a Bessel function $J_n^*(\beta)$ which has been normalized with respect to the maximum amplitude of the calculated frequency spectrum (obtained using Eq. (2)):

$$\mathcal{S}_{\mathrm{LI}} \simeq \frac{\mathcal{A} V_0}{2} \sum_{n=0}^{\lceil \beta/2 \rceil} (\mathcal{S}_{\mathrm{CW}}[\omega_c + n\nu 2\pi, |J_n^*(\beta)\Omega|]$$
$$- \mathcal{S}_{\mathrm{CW}}[\omega_c - n\nu 2\pi, |J_n^*(\beta)\Omega|]). \quad (4)$$

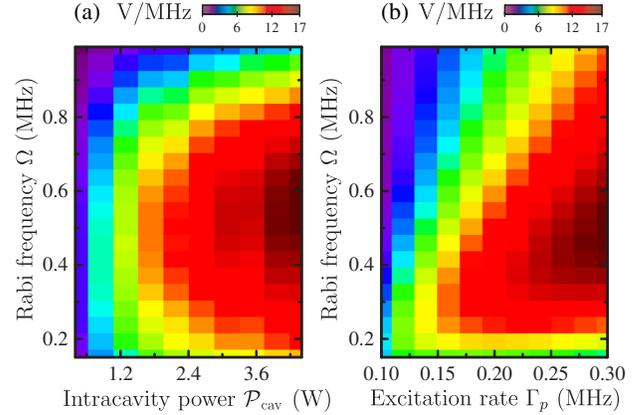

FIG. 4. (a) Measured $\max\{d_\omega \mathcal{S}_{\mathrm{LI}}\}$ for $\omega_c = \omega_0$ as a function of $\Omega$ and $\Gamma_p$. (b) Simulated comparison of the plot in (a) using Eq. (4) and identical experimental parameters of $\mathcal{A} = 5 \times 10^4$, $T_1 \sim 5.5$ ms, $T_2^* \sim 0.4$ $\mu$s, and $m = 0.5$ MHz.

A generated spectrum using Eq. (4) is fitted to a measured spectrum in Fig. 3(c), showing excellent agreement.

With an optimized $m$, $\nu$, and drive frequency composition, exploration of the trend of $\max\{d_\omega \mathcal{S}_{\mathrm{LI}}\}$ with varying $\Gamma_p$'s and $\Omega$'s is carried out for a fixed lock-in gain and time constant. The measured points are plotted in Fig. 4(a). Using the equivalent experimental lock-in gain settings (giving a prefactor of $g = 5 \times 10^3$), coherence times obtained through pulsed measurements, and decay rates inferred from Ref. [33], a simulated comparison is shown in Fig. 4(b), using Eq. (4) and the five-level model summarized in the Supplemental Material [28]. Both measured and simulated trends indicate an $\Omega$-dependent maximum, while there is a general increase with $\Gamma_p$. Specifically, the increase in $\Gamma_p$ is seen to offset the degradation of $\max\{d_\omega \mathcal{S}_{\mathrm{LI}}\}$ associated with power broadening when $\Omega$ increases. Essentially, this is the linewidth-narrowing effect [24,35], where the onset of power broadening is counteracted by the increased depletion rate of $^3A_2$. This mechanism is accentuated by adjusting the $\Gamma_p/\Omega$ ratio, which leads to obtaining an optimum $\max\{d_\omega \mathcal{S}_{\mathrm{CW}}\}$. This regime is reached by virtue of the cavity enhancement, allowing for the necessary $\Gamma_p$ for the large ensemble volume to be achieved. The $\Omega$ values used for the simulation match the experimentally verified parameters, while the excitation rate is found to obey the relationship $\Gamma_p \simeq (\varepsilon/4) \mathcal{P}_{\mathrm{cav}}$. Although the trend correspondence is not perfectly identical in this part of the parameter space, the similarity of the occurrence and position of a minimum and the slope values points towards the validity of expression (4).

A projection of the maximum-attainable sensitivity can be calculated assuming detection is limited solely by the lock-in input noise and the detected shot noise. The projected sensitivity can be estimated from the ratio of the electronic noise level and the point in the spectrum where the fastest rate of fluorescence change occurs when a magnetic field is applied, i.e., the point of maximum slope max$\{d_\omega S_{LI}\}$. Considering an ensemble of $10^9$ emitters and a collection efficiency of 2%, the shot-noise level is calculated to be approximately 58 nV/$\sqrt{\text{Hz}}$. Using the estimated lock-in input noise and detector load-generated noise (approximately 7 nV/$\sqrt{\text{Hz}}$ and 13 nV/$\sqrt{\text{Hz}}$, respectively), along with the previously stated gain setting $\mathcal{A} = 5 \times 10^4$ and the coherence times for the measurements in Fig. 4(a), a maximum projected sensitivity is calculated as

$$\delta B = \frac{\mathcal{A}(80\text{nV}/\sqrt{\text{Hz}})}{\max\{d_\omega S_{LI}\}\gamma_e} \sim 160 \text{ pT}/\sqrt{\text{Hz}} \qquad (5)$$

for $\Omega = 5.7$ MHz and $\Gamma_p = 6$ MHz, which corresponds to a cavity input power of $\mathcal{P}_{\text{in}} = 1.1$ W. While $\delta B$ should scale inversely with the square root of the number of N-$V$ centers, the measured absorption fraction and simulations indicate that the expected excitation power needed to optimize max$\{d_\omega S_{LI}\}$ is strongly nonlinear with respect to the number of emitters. While the actual increase in excitation power will depend on how exactly the N-$V$ densities are increased and the resulting ratio between N-$V$ centers and other absorbing impurities, this suggests that increasing ensemble densities to maximize sensitivities may not be an optimal strategy, as the power required to optimize max$\{d_\omega S_{LI}\}$ may be difficult to experimentally maintain. Instead, further improvement of the collection efficiencies and detection electronics (in terms of photodetector responsiveness and low-noise preamplification) would be more practically realizable.

## IV. ENHANCED MAGNETIC FIELD SENSING

Optimization of the lock-in modulation and excitation rates provides an optimally sensitive magnetic field probe, and by setting $\omega_c = \omega_0$ (the point of max$\{d_\omega S_{LI}\}$), the presence of magnetic fields is observed through the change in the detected voltage, with a scaling inversely proportional to $\gamma_e$. Experimental assessment of the resulting sensitivity is carried out through both measuring the magnetic noise spectral density, as well as generating a weak oscillating field close to the diamond, the results of which are summarized in Fig. 5.

For the measurement of the spectral noise density, groups $A$ and $B$ are both assessed. With $\omega_c = \omega_0$, a time trace of $5 \times 10^5$ samples is recorded with a 2-kHz sampling rate, and the resulting sensitivity (T/$\sqrt{\text{Hz}}$) scaling of the trace's Fourier transform is obtained. This measurement is also carried out for a MW excitation that is off resonance ($\omega_c \gg \omega_0$) in an insensitive part of the spectrum. The resulting spectra in Fig. 5(a) show distinct spectral features resting on a approximately 200-pT/$\sqrt{\text{Hz}}$ noise floor for both groups $A$ and $B$. Given the theoretically anticipated sensitivity calculated in the previous section, this value is deemed to be near-shot-noise limited for the current excitation parameters of $\Omega \sim 0.55$ MHz, and $\Gamma_p \sim 0.3$ MHz ($\mathcal{P}_{\text{in}} \sim 0.4$ W). Low frequency (dc to 5-Hz) noise is attributed to the slow, temperature-dependent, fluctuating magnetization of the permanent magnet and to other residual magnetizations of surrounding metallic components. It is also affected by the temperature dependence of the N-$V$ resonance frequency, which has a temperature-dependent zero-field splitting adding to the shifted spin resonances $\omega_0$ by a factor of $-74.2$ kHz/K [36]. Group $B$ is expected to be much more sensitive to such low-frequency magnetic noise due to the three subgroups possessing different relative projection angles, and it therefore displays a higher noise density below 5 Hz. Towards higher frequencies, there is a characteristic first-order filter roll-off ($-20$ dB/decade) which occurs after the cutoff frequency at 159 Hz. Most noteworthy is the detection of the 50-Hz magnetic mains "hum" and its subsequent odd harmonics, shown to be detectable only when the MW is on resonance. Detection of 50 Hz and its second odd harmonic for both groups are shown in Figs. 5(a)(i) and 5(a)(ii), subject to the characteristic dampening of the low-pass filter. For comparison, the noise spectra generated from the lock-in amplifier at its lowest gain setting with the connected blocked detector is also plotted, with a similar sensitivity scaling as for group $A$. This represents the cumulative noise floor limit of the current configuration, which may be overcome by further enhancing max$\{d_\omega S_{LI}\}$ (i.e., via changing the linewidth-to-contrast ratio) and increasing the light collection efficiency.

Further investigation of the system's noise characteristics is carried out by calculating the Allan deviation of the time traces used to calculate the noise spectra in Fig. 5(a). Using the color designation in Fig. 5(a), the Allan deviations are plotted in Fig. 5(b) in units of $T$. The Allan deviation is a measure of the standard deviation as a function of sample binning size and is commonly used, in conjunction with the spectral noise density, to study and identify the presence of systematic and stochastic noise in oscillatory systems [37]. The slopes and features of the obtained trend highlight the type of noise present and their limits on the optimal averaging time. In the calculated Allan plots, the most striking feature is the clear difference in behavior between the on- and off-resonance traces. The off-resonance traces exhibits a constant $\tau^{-1/2}$ scaling, which signifies the dominance of stochastic white noise, as expected from thermally induced electronic noise generated in the detector and lock-in amplifier components. Both on-resonance traces also show the same trend dressed with the systematic noise originating from the 50-Hz hum, but they reach slightly different minimum $\tau$'s of approximately 4 nT at 0.4 s and about 6 nT at 0.2 s for $A$ and $B$, respectively. Group $B$ displays a higher Allan deviation on a slightly shorter time scale, believed to be due to the increased sensitivity of three groups maintaining degeneracy, and the larger intrinsic magnetic noise inherent to the denser ensemble of group $B$ compared to group $A$. For longer averaging times, the minima are followed by an increase with a scaling that signifies the onset of long-term thermal drift, originating from, e.g., temperature-based magnetization fluctuations and the thermal expansion and displacement of surrounding metallic fixtures.

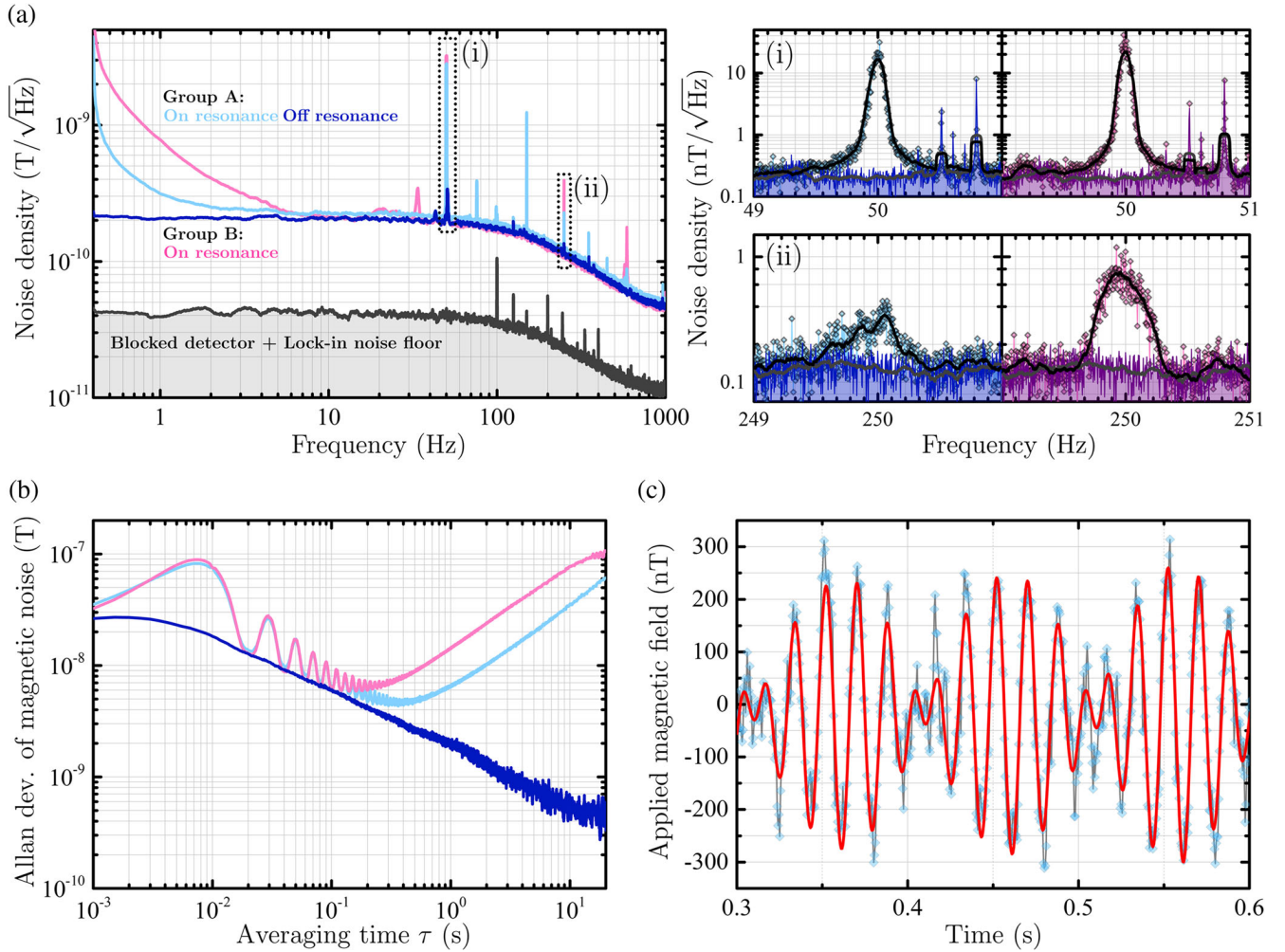

FIG. 5. (a) Plots of the magnetic noise spectral density when on resonance ($\omega_c = \omega_0$) and off resonance ($\omega_c \gg \omega_0$), for both groups $A$ and $B$ with $\Omega \sim 0.55$ MHz and $\Gamma_p \sim 0.3$ MHz ($\mathcal{P}_{in} \sim 0.4$ W). The combined noise floor of the lock-in and detector for the lowest lock-in gain settings, assuming a similar $\max\{d_\omega \mathcal{S}_{LI}\}$ to that of $A$, is also plotted. The plots are averaged from five subsequently measured frequency traces and are smoothed using a Savitzky-Golay filter. Magnified plots of the peaks designated (i) and (ii) show unsmoothed data points with the smoothed traces, highlighting the detection and lack of detection of the 50-Hz magnetic hum and its second odd harmonic when on or off resonance. The difference in amplitude are attributed to varying laboratory conditions. (b) Plots of the Allan deviation of the traces in (a) using an identical color designation. The trends show drops scaling with $\tau^{-1/2}$ highlighting the dominance of white noise in this regime, while the on-resonance plots are dressed with the systematic noise originating from the 50-Hz magnetic hum. A minimum floor for on-resonance detection is reached for $\tau = 0.2$–$0.4$ s, which signifies the limitation of electronic voltage noise ("flicker" noise), with further averaging (at large $\tau$) giving no advantage. The subsequent increase highlights the dominance of long-term drift through thermal- and mechanical-based Brownian noise. The larger Allan deviation of group $B$ is related to the fact the maintaining degeneracy is more noisy. (c) Time trace plot of the on-resonance response of group $B$ to a generated 60-Hz magnetic ac field using the coil. The beating of the 60-Hz noise and the 50-Hz magnetic hum is made clear by the red trace, which is a guide for the eye.

Finally, a weak ac magnetic field is applied to test the system response to more realistic measurement scenarios. A 60-Hz sine field is generated using the coil placed near the diamond holder, and a time trace is measured for group $B$, using all of the same measurement parameters applied for measuring the magnetic spectral noise density in Fig. 5(a). A portion of a recorded 250-s time trace is plotted in Fig. 5(c), which distinctly highlights beating between the generated 60-Hz field and the 50-Hz magnetic hum. Bearing in mind the Allan-deviation characteristics of the magnetometer shown in Fig. 5(b), estimating a single standard deviation of the measured trace would only be fair in the presence of a flat spectrum dominated by white noise. Instead, an Allan deviation of approximately 6 nT is obtained for a sampling bin of 0.2 s from the total 250-s trace [a portion of which is shown Fig. 5(c)]. Accounting for the sampling-bin bandwidth, a directly extracted sensitivity of about $3 \text{ nT}/\sqrt{\text{Hz}}$ is calculated. This value is larger then the measured noise density floor due to the presence of the low-frequency magnetic noise fluctuations from the environment and the magnetic coil itself. Through shielding of the environment and stabilization of the magnetic coils current source, this measured value is expected to approach the noise floor of approximately $200 \text{ pT}/\sqrt{\text{Hz}}$.

## V. CONCLUSION AND OUTLOOK

This work demonstrates how amplifying the pump field in combination with a microwave resonator antenna and lock-in detection can reach a near-shot-noise-limited noise floor using an off-the-shelf CVD diamond with no additional modification and unoptimized fluorescence collection. The use of the confocal cavity brings about excitation across a volume of approximately $3.5 \times 10^{-2}$ mm$^3$, with an estimated magnitude of about $10^9$ N-$V$ centers, allowing for a nonlinear fluorescence excitation regime to be reached. Reaching this excitation regime was crucial for bringing about linewidth narrowing and thereby maximizing the measured lock-in slope, in combination with simultaneously exciting all three hyperfine frequencies. The enhanced sensitivity of the signal allows for the measurement of an approximately 200-pT$\sqrt{\text{Hz}}$ noise floor for frequencies ranging from 0.1 to 159 Hz (limited by the time constant set by the lock-in amplifier), and the measurement of an applied 60-Hz magnetic field with an extracted sensitivity of about $3 \text{ nT}/\sqrt{\text{Hz}}$ which is limited by the environmental magnetic noise. Compared to the projected calculated sensitivity of approximately 160 pT$\sqrt{\text{Hz}}$ that is achievable with the estimated number of N-$V$ centers, the measured noise floor is assumed to be near-shot-noise limited.

This work shows the importance of enhancing the efficiency and uniformity of excitation and spin polarization, with measured near-dc sensitivities comparable to those reported in the literature that are based on using denser N-$V$ ensembles. Furthermore, it indicates that an increased ensemble density may necessitate significantly larger excitation powers to optimize their collective sensitivity. While this statement may be obvious, these results suggest that the needed power may be impractical from an application point of view, even if N-$V$ absorption is the dominant absorption source.

## ACKNOWLEDGMENTS

We would like to thank Kristian Hagsted Rasmussen for help with diamond surface preparation and Ilya Radko for supporting N-$V$ concentration measurements. We are also thankful to Fedor Jelezko, Adam Wojciechowski, and Ilja Gerhardt for helpful discussions. This work is partly funded by Innovation Fund Denmark under the EXMAD project and the Qubiz center, as well as the Danish research Council under the DIMS project.

S. A. and H. A. R. E.-E. contributed equally to this work.

[1] L. Rondin, J. P. Tetienne, T. Hingant, J. F. Roch, P. Maletinsky, and V. Jacques, Magnetometry with nitrogen-vacancy defects in diamond, Rep. Prog. Phys. **77**, 056503 (2014).

[2] G. Kucsko, P. C. Maurer, N. Y. Yao, M. Kubo, H. J. Noh, P. K. Lo, H. Park, and M. D. Lukin, Nanometre-scale thermometry in a living cell, Nature (London) **500**, 54 (2013).

[3] P. Neumann, I. Jakobi, F. Dolde, C. Burk, R. Reuter, G. Waldherr, J. Honert, T. Wolf, A. Brunner, J. H. Shim, D. Suter, H. Sumiya, J. Isoya, and J. Wrachtrup, High-precision nanoscale temperature sensing using single defects in diamond, Nano Lett. **13**, 2738 (2013).

[4] F. Dolde, H. Fedder, M. W. Doherty, T. Nöbauer, F. Rempp, G. Balasubramanian, T. Wolf, F. Reinhard, L. C. L. Hollenberg, F. Jelezko, and J. Wrachtrup, Electric-field sensing using single diamond spins, Nat. Phys. **7**, 459 (2011).

[5] D. R. Glenn, K. Lee, H. Park, R. Weissleder, A. Yacoby, M. D. Lukin, H. Lee, R. L. Walsworth, and C. B. Connolly, Single-cell magnetic imaging using a quantum diamond microscope, Nat. Methods **12**, 736 (2015).

[6] G. Balasubramanian, I. Y. Chan, R. Kolesov, M. Al-Hmoud, J. Tisler, C. Shin, C. Kim, A. Wojcik, P. R. Hemmer, A. Krueger, T. Hanke, A. Leitenstorfer, R. Bratschitsch, F. Jelezko, and J. Wrachtrup, Nanoscale imaging magnetometry with diamond spins under ambient conditions, Nature (London) **455**, 648 (2008).

[7] L. T. Hall, G. C. G. Beart, E. A. Thomas, D. A. Simpson, L. P. McGuinness, J. H. Cole, J. H. Manton, R. E. Scholten, F. Jelezko, J. Wrachtrup, S. Petrou, and L. C. L. Hollenberg, High spatial and temporal resolution wide-field imaging of neuron activity using quantum NV-diamond, Sci. Rep. **2**, 401 (2012).

[8] J. M. Taylor, P. Cappellaro, L. Childress, L. Jiang, D. Budker, P. R. Hemmer, A. Yacoby, R. Walsworth, and M. D. Lukin, High-sensitivity diamond magnetometer with nanoscale resolution, Nat. Phys. **4**, 810 (2008).

[9] Romana Schirhagl, Kevin Chang, Michael Loretz, and Christian L. Degen, Nitrogen-vacancy centers in diamond: Nanoscale sensors for physics and biology, Annu. Rev. Phys. Chem. **65**, 83 (2014).

[10] N. Y. Yao, L. Jiang, A. V. Gorshkov, P. C. Maurer, G. Giedke, J. I. Cirac, and M. D. Lukin, Scalable architecture for a room temperature solid-state quantum information processor, Nat. Commun. **3**, 800 (2012).

[11] D. Arnold, S. Siegel, E. Grisanti, J. Wrachtrup, and I. Gerhardt, A rubidium M$_x$-magnetometer for measurements on solid state spins, Rev. Sci. Instrum. **88**, 023103 (2017).

[12] J. F. Barry, M. J. Turner, J. M. Schloss, D. R. Glenn, Y. Song, M. D. Lukin, H. Park, and R. L. Walsworth, Optical magnetic detection of single-neuron action potentials using quantum defects in diamond, Proc. Natl. Acad. Sci. U.S.A. **113**, 14133 (2016).

[13] H. Clevenson, M. E. Trusheim, T. Schroder, C. Teale, D. Braje, and D. Englund, Broadband magnetometry and temperature sensing with a light trapping diamond waveguide, Nat. Phys. **11**, 393 (2015).

[14] D. Le Sage, L. M. Pham, N. Bar-Gill, C. Belthangady, M. D. Lukin, A. Yacoby, and R. L. Walsworth, Efficient photon detection from color centers in a diamond optical waveguide, Phys. Rev. B **85**, 121202 (2012).

[15] T. Wolf, P. Neumann, K. Nakamura, H. Sumiya, T. Ohshima, J. Isoya, and J. Wrachtrup, Subpicotesla Diamond Magnetometry, Phys. Rev. X **5**, 041001 (2015).

[16] V. M. Acosta, E. Bauch, M. P. Ledbetter, C. Santori, K.-M. C. Fu, P. E. Barclay, R. G. Beausoleil, H. Linget, J. F. Roch, F. Treussart, S. Chemerisov, W. Gawlik, and D. Budker, Diamonds with a high density of nitrogen-vacancy centers for magnetometry applications, Phys. Rev. B **80**, 115202 (2009).

[17] G. Balasubramanian, P. Neumann, D. Twitchen, M. Markham, R. Kolesov, N. Mizuochi, J. Isoya, J. Achard, J. Beck, J. Tissler, V. Jacques, P. R. Hemmer, F. Jelezko, and J. Wrachtrup, Ultralong spin coherence time in isotopically engineered diamond, Nat. Mater. **8**, 383 (2009).

[18] P. Siyushev, F. Kaiser, V. Jacques, I. Gerhardt, S. Bischof, H. Fedder, J. Dodson, M. Markham, D. Twitchen, F. Jelezko, and J. Wrachtrup, Monolithic diamond optics for single photon detection, Appl. Phys. Lett. **97**, 241902 (2010).

[19] J. P. Hadden, J. P. Harrison, A. C. Stanley-Clarke, L. Marseglia, Y. L. D. Ho, B. R. Patton, J. L. O'Brien, and J. G. Rarity, Strongly enhanced photon collection from diamond defect centers under microfabricated integrated solid immersion lenses, Appl. Phys. Lett. **97**, 241901 (2010).

[20] B. J. M. Hausmann, B. J. Shields, Q. Quan, Y. Chu, N. P. De Leon, R. Evans, M. J. Burek, A. S. Zibrov, M. Markham, D. J. Twitchen, H. Park, M. D. Lukin, and M. Loncr, Coupling of NV centers to photonic crystal nanobeams in diamond, Nano Lett. **13**, 5791 (2013).

[21] D. Riedel, D. Rohner, M. Ganzhorn, T. Kaldewey, P. Appel, E. Neu, R. J. Warburton, and P. Maletinsky, Low-Loss Broadband Antenna for Efficient Photon Collection from a Coherent Spin in Diamond, Phys. Rev. Applied **2**, 064011 (2014).

[22] Y. Dumeige, M. Chipaux, V. Jacques, F. Treussart, J. F. Roch, T. Debuisschert, V. M. Acosta, A. Jarmola, K. Jensen, P. Kehayias, and D. Budker, Magnetometry with nitrogen-vacancy ensembles in diamond based on infrared absorption in a doubly resonant optical cavity, Phys. Rev. B **87**, 155202 (2013).

[23] K. Jensen, N. Leefer, A. Jarmola, Y. Dumeige, V. M. Acosta, P. Kehayias, B. Patton, and D. Budker, Cavity-Enhanced Room-Temperature Magnetometry Using Absorption by Nitrogen-Vacancy Centers in Diamond, Phys. Rev. Lett. **112**, 160802 (2014).

[24] K. Jensen, V. M. Acosta, A. Jarmola, and D. Budker, Light narrowing of magnetic resonances in ensembles of nitrogen-vacancy centers in diamond, Phys. Rev. B **87**, 014115 (2013).

[25] N. Aslam, G. Waldherr, P. Neumann, F. Jelezko, and J. Wrachtrup, Photo-induced ionization dynamics of the nitrogen vacancy defect in diamond investigated by single-shot charge state detection, New J. Phys. **15**, 013064 (2013).

[26] R. U. A. Khan, B. L. Cann, P. M. Martineau, J. Samartseva, J. J. P. Freeth, S. J. Sibley, C. B. Hartland, M. E. Newton, H. K. Dhillon, and D. J. Twitchen, Colour-causing defects and their related optoelectronic transitions in single crystal CVD diamond, J. Phys. Condens. Matter **25**, 275801 (2013).

[27] T.-L. Wee, Y.-K. Tzeng, C.-C. Han, H.-C. Chang, W. Fann, J.-H. Hsu, K.-M. Chen, and Y.-C. Yu, Two-photon excited fluorescence of nitrogen-vacancy centers in proton-irradiated type Ib diamond, J. Phys. Chem. A **111**, 9379 (2007).

[28] See Supplemental Material at http://link.aps.org/supplemental/10.1103/PhysRevApplied.8.034001 for a wide-span measurement of the mixed MW signal, a description of the Bloch equations, its steady-state solution, and a derivation of the Fourier series for a frequency-modulated wave using a sine function.

[29] P. L. Stanwix, L. M. Pham, J. R. Maze, D. Le Sage, T. K. Yeung, P. Cappellaro, P. R. Hemmer, A. Yacoby, M. D. Lukin, and R. L. Walsworth, Coherence of nitrogen-vacancy electronic spin ensembles in diamond, Phys. Rev. B **82**, 201201(R) (2010).

[30] R. Fischer, A. Jarmola, P. Kehayias, and D. Budker, Optical polarization of nuclear ensembles in diamond, Phys. Rev. B **87**, 125207 (2013).

[31] A. Wickenbrock, H. Zheng, L. Bougas, N. Leefer, S. Afach, A. Jarmola, V. M. Acosta, and D. Budker, Microwave-free magnetometry with nitrogen-vacancy centers in diamond, Appl. Phys. Lett. **109**, 053505 (2016).

[32] H. A. R. El-Ella, S. Ahmadi, A. Wojciechowski, A. Huck, and U. L. Andersen, Optimised frequency modulation for continuous-wave optical magnetic resonance sensing using nitrogen-vacancy ensembles, Opt. Express **25**, 14809 (2017).

[33] L. Robledo, H. Bernien, T. van der Sar, and R. Hanson, Spin dynamics in the optical cycle of single nitrogen-vacancy centres in diamond, New J. Phys. **13**, 025013 (2011).

[34] J. R. Carson, Notes on the theory of modulation, Proc. IRE **10**, 57 (1922).

[35] A. Dréau, M. Lesik, L. Rondin, P. Spinicelli, O. Arcizet, J.-F. Roch, and V. Jacques, Avoiding power broadening in optically detected magnetic resonance of single NV defects for enhanced dc magnetic field sensitivity, Phys. Rev. B **84**, 195204 (2011).

[36] V. M. Acosta, E. Bauch, M. P. Ledbetter, A. Waxman, L.-S. Bouchard, and D. Budker, Temperature Dependence of the Nitrogen-Vacancy Magnetic Resonance in Diamond, Phys. Rev. Lett. **104**, 070801 (2010).

[37] D. W. Allan, Should the classical variance be used as a basic measure in standards metrology?, IEEE Trans. Instrum. Meas. **IM-36**, 646 (1987).